\documentclass[12pt]{article}
\usepackage{latexsym}
\usepackage{amsmath}
\usepackage{amssymb}
\begin{document}
\begin{center}
\Large
\textbf{Remarks on the Relativistic
Transactional Interpretation of
Quantum Mechanics}\\[0.5cm]
\large
\textbf{Louis Marchildon}\\[0.5cm]
\normalsize
D\'{e}partement de chimie, biochimie et physique,\\
Universit\'{e} du Qu\'{e}bec,
Trois-Rivi\`{e}res, Qc.\ Canada G9A 5H7\\[0.2cm]
(louis.marchildon$\hspace{0.3em}a\hspace{-0.8em}\bigcirc$uqtr.ca)\\
\end{center}
%
%
\begin{abstract}
Kastner (arXiv:1709.09367) and Kastner and
Cramer (arXiv:1711. 04501) argue that the
Relativistic Transactional Interpretation
(RTI) of quantum mechanics provides a clear
definition of absorbers and a solution
to the measurement problem. I briefly examine
how RTI stands with respect to unitarity
in quantum mechanics. I then argue that a
specific proposal to locate the origin of
nonunitarity is flawed, at least in its
present form.
\end{abstract}
\section{Introduction}
It is generally agreed that the state
vector of a nonrelativistic quantum system
normally evolves unitarily, according to
the Schr\"{o}dinger equation. A number of
investigators think that `normally'
extends to `universally.' That is, they
believe that unitary evolution suffers no
exceptions, no matter how large the system.
Others, following von Neumann, believe that
unitary evolution breaks down in some
circumstances, measurements in particular.
In such situations, so they think, the
Schr\"{o}dinger equation must be replaced
by what has been known as the collapse of
the state vector. Although von Neumann did
not specify the mechanism of collapse, other
investigators after him attempted to do so,
for instance in approaches like spontaneous
localization or gravitational collapse. By
contrast, the approaches initiated by de
Broglie and Bohm, and by Everett, stick to
unitarity.

Cramer's Transactional Interpretation (TI)
of quantum mechanics~\cite{cramer1,cramer2}
stands on the collapse side. Cramer's views
can be illustrated by the process of emission
and absorption of a quantum particle.
Associated with emission is an `offer wave'
that travels forward in time. This wave
reaches a number of possible absorbers.
Each of these responds with a `confirmation
wave' that travels backward in time to the
emitter. A transaction is henceforth
established between the emitter and one of
the absorbers, whose probability is naturally
given by the Born rule. A transaction is taken
to be irreversible and, in the end, it
corresponds to state vector collapse.

In a review~\cite{marchildon} of {\it The
Quantum Handshake}~\cite{cramer2}, I argued
that although TI sheds light on some
paradoxical aspects of quantum mechanics,
it is not really defined better than von
Neumann's collapse or, indeed, than the
Copenhagen interpretation. This is essentially
because transactions require emitters and
absorbers, which share the ambiguity of
Bohr's classical objects or von Neuman's
measurement apparatus. TI therefore
reintroduces, under a different guise, the
quantum-classical distinction.

In reply to~\cite{marchildon},
Kastner~\cite{kastner1} and Kastner and
Cramer~\cite{kc} have argued that in the
Relativistic Transactional Interpretation
(RTI) developed by Kastner~\cite{kastner2},
absorbers are indeed well-defined. Moreover,
they claim that RTI provides a solution
to the measurement problem. In this note
I put these claims in perspective and argue
that they are overstated.
\section{Nonunitarity in RTI}
As explained in~\cite{kastner2} and
elsewhere (see~\cite{kastner1} for
references), RTI is based on the Davies
quantum relativistic direct-action theory.
I shall not review the full theory here,
but focus on what is relevant to the
points I want to make.

A crucial difference between RTI and TI
is that in the former, absorbers do not
need to be macroscopic. To quote Kastner,
``Emission and absorption are [in RTI]
quantitatively defined at the microscopic
level'' \cite[p.~2]{kastner1}. Confirmation
waves, which trigger absorption, are
defined at the level of interacting quantum
fields: ``the coupling amplitudes between
interacting fields in the relativistic
domain are to be identified as the
amplitudes for the generation of confirmation
waves'' \cite[p.~65]{kastner2}.

Emission and absorption are
associated with offer and confirmation waves,
respectively. Kastner~\cite[p.~4]{kastner1}
stresses that ``absorption \mbox{[\ldots]}
{\it is irreversible (non-unitary) at the
level of the micro-absorber}.'' Moroever,
she proposes a quantitative measure for
the generation of an offer or confirmation
wave~\cite[p.~5]{kastner1}:
\begin{quote}
The crucial development allowing definition
of measurement in the relativistic RTI is
[that the] coupling amplitude
$e$ ({\it natural units}) is identified as
{\it the amplitude for an offer or 
confirmation to be generated}.
\end{quote}
For an interaction between a photon and an
electron, the coupling amplitude is equal
to the electron charge~$e$. The probability
of generation of an offer or a confirmation
wave in such a microscopic process is then
equal to the square of the amplitude, that is,
to the fine structure constant $\alpha \approx
1/137$.\footnote{In~\cite{kastner2,kastner3}
Kastner associates $\alpha$ with the amplitude
and $\alpha^2$ with the probability. The
forthcoming arguments can be adapted to
that specification.}

Thus in RTI, the origin of nonunitarity
lies in the microscopic world. But the
evidence for nonunitarity (for those who,
unlike Bohmians and Everettians, believe
in it) comes from the macroscopic world,
that is, from the apparently well-defined
results of measurements. Does the
microscopic nonunitarity explain the
macroscopic nonunitarity? Perhaps, but this
is far from obvious.\footnote{The logical
gap is not unlike the one between quantum
indeterminacy and human free will. In the
absence of proof, the claim that the former
explains the latter is just a hypothesis.
(To avoid any misunderstanding, I should
say that I do not attribute such
a claim to Kastner or Cramer.)} An
explanation of the one by the other should
be buttressed by detailed calculations.
I am not aware of any such calculations
performed in the framework of RTI but,
fortunately, estimates have been made.
\section{Quantitative Estimates}
According to~\cite{kastner1}, the
probability of generation of a confirmation
wave by a single absorber of charge~$e$ is
small, on the order of $\alpha \approx
0.0073$. This probability, however, will
increase with the number of absorbers. The
argument goes this way~\cite{kastner1}.
Suppose that an offer wave is emitted,
and that there are $N$~possible absorbers.
The probability that any of these will not
generate a confirmation wave is equal to
$1 - \alpha \approx 0.9927$. Assuming that
absorbers are independent, the probability
that none of them will generate a
confirmation wave is therefore on
the order of $0.9927^N$, a very small
number if $N$ is large. Hence
macroscopic absorbers will generate
confirmation waves, and therefore
nonunitarity and irreversibility,
essentially with certainty.\footnote{Such an
argument has also been used in spontaneous
localization theories, where the localization
of single particles occurs very infrequently
but the localization of macroscopic objects
occurs very quickly. This remark should not
be construed as identifying RTI with
spontaneous localization theories.}

This argument also allows estimating the
values of~$N$ for which absorption occurs with
neither too small nor too large probabilities.
Kastner points out that if, for example,
each of the 60~carbon atoms of a buckeyball
acts as a microscopic absorber, we should
expect absorption in about one third of cases
($1 - 0.9927^{60} \approx 0.36$).
This means that systems like buckeyballs
should exhibit nonunitarity and irreversibility
in a substantial proportion of cases. Of course,
a full argument will need to be more specific,
and consider detailed experimental circumstances
in which the buckeyball is investigated.
Interestingly, however, buckeyballs have
already been shown to display quantum
interference~\cite{arndt}.
This implies that in such situations,
their behavior is consistent with unitarity
and reversibility. Advocates of RTI will point
out that buckeyball interference does not
involve absorption, so that unitarity should
be expected. But the point is that we already
have the technology to work with a number of
atoms for which, under appropriate
circumstances, RTI would predict a breakdown
of unitarity. If indeed $\alpha$ represents the
probability of generation of confirmation
waves in microscopic RTI, we may soon be able
to distinguish predictions made by unitary
quantum mechanics from those made by
nonunitary theories like RTI.\footnote{Some
spontaneous localization theories have
already been ruled out in similar contexts.}
\section{A problem with Probability}
In a process involving a photon and an
electron, the probability of generation
of offer and confirmation waves is equal
to $e^2$, the fine structure constant.
Kastner further asserts that in this
context $\alpha \approx
1/137$ represents the probability of a
measurement transition, that is, the
probability of an actual physical process.
I will now argue that such an identification
is unlikely to hold as any general
principle.

The argument goes this way. Suppose there were
in nature fundamental particles with charge
$12 e$. The probability that they would
generate confirmation waves, according to
Kastner's prescription, should be equal to
$(12 e)^2 \approx 144/137 \approx 1.05$.
That is, the probability would be larger
than one. Clearly, this cannot be interpreted
as the probability of a physical process,
or as any probability whatsoever.
One can reply that there are no fundamental
particles with charge $12 e$ but, since this
is a contingent rather than necessary fact,
the reply is unconvincing. Moreover, Kastner
argues~\cite{kastner1} that her prescription
generalizes to other kinds of charge, like
color. But these charges, in some energy
ranges and in natural units, have values that
exceed one. The amplitude of generation of a
confirmation wave associated with such
quantized fields cannot be equal to the charge.

Can one raise a similar objection to
standard quantum electrodynamics? The
answer is no, because in QED the charge
is not associated with the amplitude of
a physical process. The charge is
associated with vertices of Feynman
diagrams which, for a variety of reasons
(not the least of which being that
diagrams are gauge dependent), do not
individually have physical meaning.
Only the absolute square of the series
of Feynman diagrams represents a physical
process and has meaning. Should there be
fundamental particles of charge $12 e$,
they would not lead to probabilities greater
than one. They would just make the series
useless.
\section{Conclusion}
I have shown that the quantitative
criterion proposed by Kastner for the
generation of offer and confirmation
waves faces serious difficulties.
This leaves the measurement problem
unsolved and absorbers ill-defined.
Comparing predictions of nonunitary TI (or
RTI) with those of unitary quantum mechanics
would be highly interesting, but this
can only be achieved through a quantitative
and consistent approach to absorption.
\section*{Acknowledgements}
Although I do not expect Ruth Kastner to
agree with my conclusions, I thank her for
email exchanges that helped sharpening my
views.
\end{document}